# High Throughput Screening of Expression Constructs using Microcapillary Arrays

Khushank Singhal[1], Harry E. Adamson[2], Thomas M. Baer[3], Howard M. Salis[2,4], Melik C. Demirel[1,5,*]

[1] Department of Engineering Science and Mechanics, The Pennsylvania State University, University Park, Pennsylvania 16802, USA.

[2] Department of Chemical Engineering, The Pennsylvania State University, University Park, Pennsylvania 16802, USA.

[3] Stanford Photonics Research Center, Stanford University, Stanford, California, USA.

[4] Department of Agricultural and Biological Engineering, The Pennsylvania State University, University Park, Pennsylvania 16802, USA.

[5] Huck Institutes of Life Sciences and Materials Research Institute, The Pennsylvania State University, University Park, Pennsylvania 16802, USA.

*Corresponding author, email: melik@psu.edu

**Keywords:** High-throughput screening, biomanufacturing, proteins, expression constructs, plasmid libraries

**ABSTRACT**

Gene expression is a complex phenomenon involving numerous interlinked variables, and studying these variables to control expression is essential in bioengineering and biomanufacturing. While cloning techniques for achieving plasmid libraries that cover large design spaces exist, multiplex techniques offering cell culture screening at similar scales are still lacking. We introduced a microcapillary array-based platform aimed at high-throughput, multiplex screening of miniature cell cultures through fluorescent reporters. The clone recovery mechanism provides 100x enrichment ratios compared to traditional techniques for establishing phenotype-to-genotype linkages. We conducted experiments to delineate the effects of three key plasmid design features—promoters, 5' untranslated regions, and amino acid sequences—on protein titer. We identified a small set of promoters that maximize protein titer from thousands of promoters with widely varying transcription rates. We established that mRNA half-lives, controlled by 5' untranslated regions, correlate with protein expression. Using dual-reporter imaging, we demonstrate relative analyses of multiple ribosome binding sites in operons. Lastly, we discuss the effect of structural protein hydrophobicity scores on their expression and cell growth profiles. Through multiple experiments with libraries of plasmid constructs, we demonstrate population binning, dual-reporter operon screening, chemical perturbation, and cell growth estimation using brightfield absorbance measurements with the platform.

Cell sorting and screening techniques are essential for biological research, bioengineering, and medicine as they facilitate the isolation of specific clones from heterogeneous populations and enable the analysis of numerous cell phenotypes in studies related to drug and immune responses[1–3], disease progression[4], enzyme and biomarker engineering[5], and cell differentiation[6]. Beyond biology and medicine, screening techniques have significant potential in the biomanufacturing and bioprocessing industries, such as biopharmaceuticals, biofuels, food and agriculture, and bioremediation, which focus on protein-titer optimization[7–9]. The key screening modality sought for biomanufacturing is population-based screening, where miniature cell cultures of individual clones can be grown and analyzed simultaneously. While significant strides in single-cell screening have been achieved using microfluidic platforms[10], progress in population-based screening remains inadequate. Microwell plate readers[11–14] and multi-parallel bioreactors[15,16] serve as essential tools for screening cell cultures in laboratories globally. However, these platforms are not ideal due to inefficient methods of analyzing large clonal libraries, insufficient throughput, or complex handling procedures. In this context, we concentrate on a population-based screening platform and illustrate its potential in biomanufacturing and gene expression studies.

Cost-effective protocols for the rapid synthesis of large plasmid libraries are being developed, and techniques that can complement these libraries with multiplex screening at similar scales are essential. Cloning strategies such as mutagenesis and combinatorial fragment assembly facilitate the construction of libraries that cover vast design spaces, which can inform protein structure-expression relationships and aid in discovering new protein sequences (e.g., those with similar properties but improved expression) through screening. Furthermore, algorithms for the massively parallel design of plasmid elements (e.g., promoters, ribosome binding sites, RBS) have opened new avenues for extensive optimization of recombinant gene expression. While predicting expression and cell growth remains complex due to astronomical cellular interdependencies[17–19], and until machine learning models can fully support these predictions[20–22], high-throughput population-based screening techniques are of substantial importance in synthetic biology (Fig. 1a). Chen and Lim et al.[23] developed a microcapillary array-based technique for high-throughput protein analysis and engineering. Through directed evolution of clones and successive screening rounds on the platform, the authors discovered a new antibody, a fluorescent protein biosensor, and an enzyme with enhanced resistance to inhibitors. The technique involves the spatial isolation of clones (>10$^5$) in capillaries and the observation of cell phenotypes over extended periods using fluorescent reporters. The platform facilitates the proliferation of cells into isogenic miniature cultures and supports a population-based screening method, unlike flow cytometry or flow-assisted cell sorting (FACS). The functional activity of designed clones is indicated by fluorescence intensity, and the clones demonstrating the highest functionality can be quickly recovered using non-contact laser-based methods extraction.

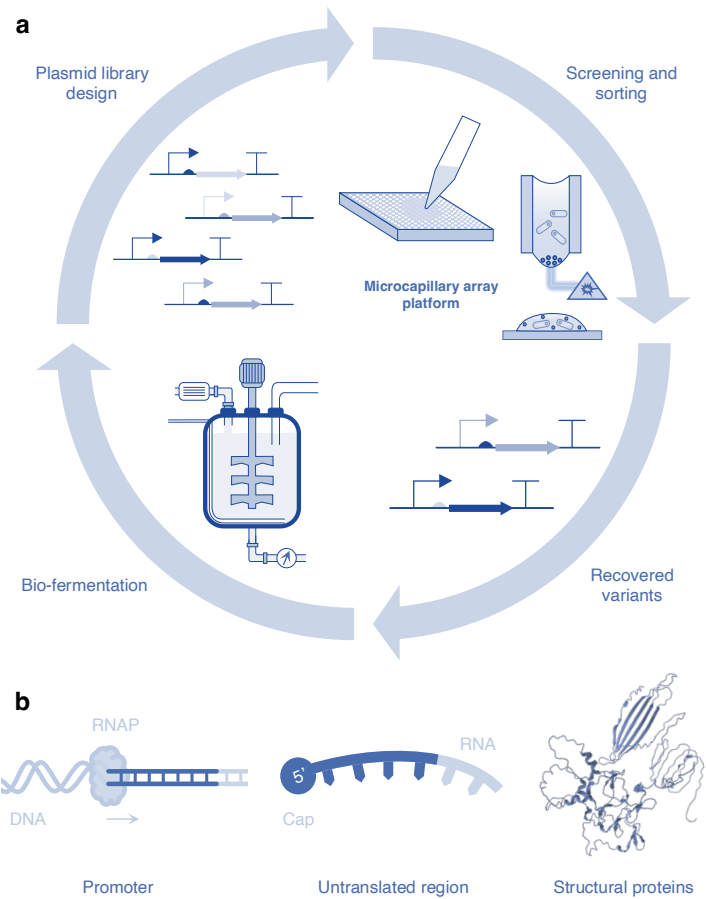

**Fig. 1. Significance of microcapillary array platform for screening expression constructs. a.** The envisioned workflow optimizes bio-manufacturability by screening and sorting plasmid libraries of expression elements. After recovering the desired clones through the platform, they can undergo traditional bio-fermentation techniques to validate titer levels and production. Insights gained from this learning step will further aid in generating new plasmid libraries. **b.** The platform's potential is demonstrated in this study by screening promoters, 5' untranslated regions, and structural protein sequences.

In this study, we developed a screening platform based on microcapillary arrays and introduced relevant features related to gene expression analysis and optimization. We incorporated a clonal binning feature to reveal biological correlations between gene expression and plasmid design rules. Our platform supports fluorophores across the visible spectrum without requiring optical modifications or excitation-wavelength switching during experiments, enabling concurrent screening of multiple fluorophores. The versatile image acquisition methods not only facilitate the screening of single-reporter plasmids but also allow for the comparative examination of multiple fluorophores with operons. Additionally, in-experiment chemical perturbation of cell cultures is enabled for studying expression phenotypes versus growth conditions (e.g., inducers[24]). Alongside fluorescence imaging, we demonstrate the assessment of cell growth profiles for individual clones through brightfield absorbance measurement[25]. This feature is particularly important for quantifying stochastic variability in gene expression and growth with toxic coding sequences (CDSs) to mitigate noise from biological factors[26–28]. The high-throughput platform maintains the laser-based mechanism for clone recovery with an enrichment ratio of up to $1:10^5$. We utilize libraries of promoters and 5' untranslated regions (UTRs) to define the impact of transcription rates, translation initiation rates, and mRNA stability on protein titer (Fig. 1b). Additionally, by employing the operonic structure of the 5' UTR library, we compare the effect of individual RBSs on the simultaneous expression of two fluorophores (mRFP1 and sfGFP). We introduce a fluorescent biosensor design to enable the quantification of non-fluorescent structural proteins (SPs) in vivo. Finally, we investigate stochastic variability in protein expression and cell growth with two structural proteins of varying hydrophobicity under different growth conditions.

## RESULTS

### Microcapillary Array-based Cell Screening

The screening methodology and the instrument are illustrated in Fig. 2. The methodology involves the spatial isolation and proliferation of cells within a lattice of micrometer-sized capillaries ($>10^5$). The open-faced capillaries maintain the cells suspended in growth media due to surface tension. The expression characteristics of constructs are monitored using fluorescent reporters, supporting both single-reporter plasmids and multi-reporter operons. The platform features a multiband illumination system accommodating fluorophores across the visible spectrum. Fluorescence signals from the capillaries are recorded and compiled automatically, and according to the screening motivation, clones exhibiting desired fluorescence characteristics are recovered to establish phenotype-to-genotype linkages. Heterogeneous mixtures of clones are loaded directly onto the arrays without prior selection or isolation. Additionally, growth characteristics are quantified through brightfield absorbance measurements across populations, particularly benefiting single-clone studies for variability.

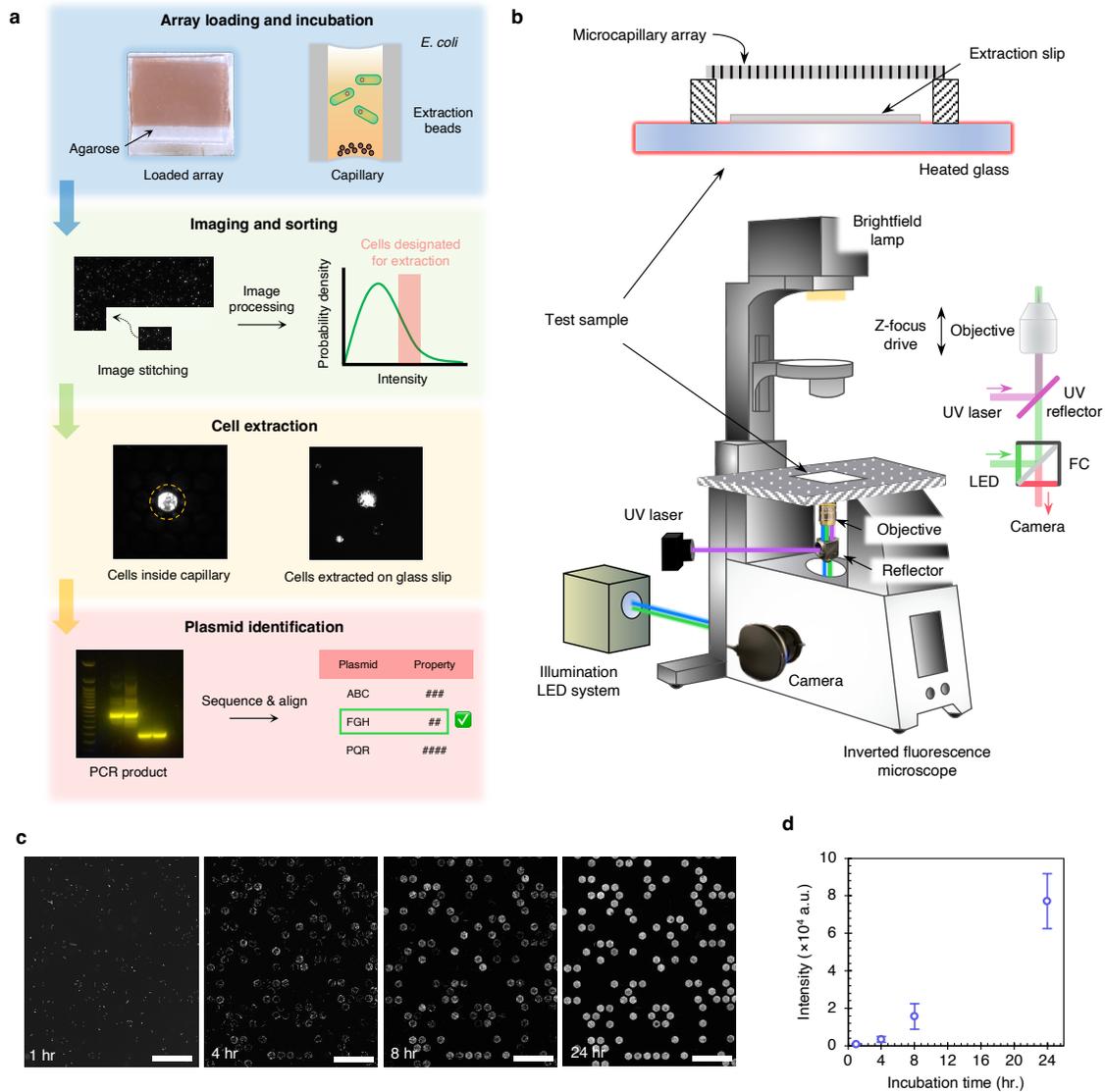

**Fig. 2. Microcapillary array screening method. a.** The methodology of the screening process is shown. The array is loaded with a mixture of magnetic beads and cells in culture media. The schematic of the cross-section of a capillary is shown. After the incubation period, snapshots of the array are acquired and stitched together. Image-processing algorithm estimates fluorescence intensities in capillaries, constructs intensity maps, and outputs coordinates of the capillaries of interest. Laser-based extraction is performed on single isolated capillaries, and the contents are collected on the extraction slip. Construct IDs on plasmids in the recovered cells are amplified through PCR, and clones are identified through sequencing and alignment. Brightness and contrast of images enhanced for visualization. **b.** The experimental setup for the screening technique is depicted. The inverted fluorescence microscope is equipped with a broad-spectrum fluorescence illumination system, filter cubes (FC), digital camera, motorized stage, and focus drive. Collimated UV laser beam is focused to a ≈ 10 μm spot in the imaging plane by the objective. The setup of the array and extraction slip on sample stage is shown. Extraction slips can be replaced during an experiment. **c.** Representative images of the array loaded with the mRFP1-expressing pFTV1 plasmid in *E. coli* after different incubation periods. The images were acquired at auto-exposure setting for better visualization of cell population within capillaries. Over time, the cells replicate and occupy the full face of the capillaries. Scale bars = 100 μm. **d.** Average capillary intensity (total capillaries = 2,695) at all incubation times.

The laser-based recovery technique precisely extracts cells from individual capillaries with each iteration (Supplementary Fig. 1a). The process uses a focused pulsed-laser beam to heat magnetic beads (for less than 20 ms) that settle at the bottom meniscus of the capillaries, creating cavitation that disrupts the meniscus for rapid release (Supplementary Fig. 1b). After extraction, the plasmids within the extracted cells are identified. Our strategy involved PCR amplification of constructs, sequencing, and alignment with reference sequences to identify the recovered clones. We designed vector-specific primer pairs for each library and collectively amplified all constructs from each extraction slip in a single PCR. The amplified constructs were then sequenced (Oxford Nanopore) and aligned. Combined with the "Context Aware Auto-align" algorithm (denovodna.com) for matching thousands of sequenced templates to references in seconds, this comprehensive protocol enables quick, accurate, and efficient identification of recovered clones.

To demonstrate cell growth and protein expression in an array, we cultured an isogenic E. coli colony that contained the plasmid expressing the red fluorescent protein mRFP1 (pFTV1). Figure 2c shows the fluorescence snapshots of the array after gradual incubation for 24 hours at 37°C. Cell growth and fluorescent protein expression are evident in these images (refer to Supplementary Figure 2a for images taken at constant exposure). In Figure 2d, the average intensity per capillary displays a non-linear increasing pattern throughout the incubation period (raw data are illustrated in Supplementary Figure 2b), likely due to the initial exponential growth phase. Notably, the inter-capillary variability in intensity is low (i.e., low standard deviation). The homogeneity of capillaries across the array ensures consistent cell growth[23].

**Transcription Rate Effects on Protein Expression in Promoter Library**

Gene promoters are key design components of plasmids that regulate protein expression by modulating the rate of DNA transcription into mRNA. We screened a library of 4,351 promoters in E. coli to categorize them based on the respective mRFP1 fluorescent reporter titers (Fig. 3a). The transcription rates of the promoters vary from 0.006 a.u. to 4731.420 a.u. (detailed sequences published earlier[29]), quantified using DNA and RNA sequencing read counts[29]. The strength of the RBS, indicated by its translation initiation rate (TIR), remains constant, with the transcription rate being the only variable controlled in the plasmids within the library.

In Fig. 3b, the fluorescence image of a section of an array is shown after a 12-hour incubation at 37°C. The image displays varying levels of mRFP1 expression, indicated by the presence of both dim and bright capillaries. We conducted three replicates of cell growth and screening with the library, and the normalized intensity maps for all replicates are depicted in Fig. 3c (see Supplementary Fig. 3 for raw intensity distributions). The trend of fluorescence intensity trends confirms consistent cell growth profiles across replicates. We extracted and identified the five best-performing promoters from the top bins in the replicates, revealing low transcription rates ranging from 5 a.u. to 15 a.u. (Fig. 3d). Compared to several stronger promoters (Supplementary Data 1), DNA-sequencing read counts for the extracted promoters were higher, while RNA-sequencing read counts were similar. This suggests that promoters associated with high mRFP1 fluorescence levels while maintaining a faster growth rate are preferentially selected. Strong promoters exhibiting elevated protein expression per cell may not be optimal for protein production due to metabolic burden or toxicity[17,18,30]. Screening the best-performing promoters also serves as a rigorous test of the platform's reproducibility. In addition to the transcription rates of the extracted promoters being similar, two of the promoters identified by replicates 1 and 3 were the same (Supplementary Data 1).

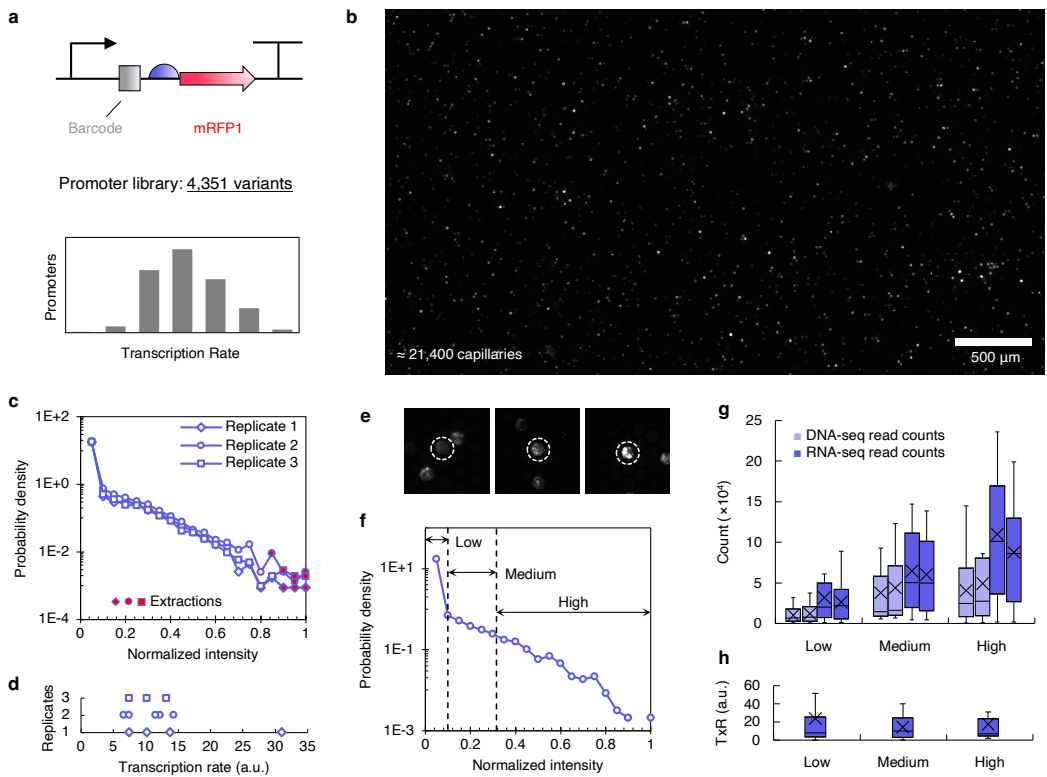

**Fig. 3. Screening of promoter library. a.** Construction of the mRFP1-expressing plasmid is shown. The library consists of 4,351 promoters with varying transcription rates, with each assigned a unique barcode. **b.** Representative image of an array loaded with the promoter library after 12 hr. incubation period. **c.** Intensity maps of three replicates of the library screening (total number of capillaries scanned in replicates 1, 2 and 3 are about 68,000, 77,000 and 72,000, resp.). Five capillaries each were extracted from the top bins. **d.** Transcription rates, estimated using the DNA/RNA-sequencing counts, of the extracted variants are plotted. Only the sequences identified with a 100% match are included in the analysis. **e.** Representative images of a capillary from the low (left, norm. intensity: original = 0.0332 a.u., after dilation = 0.05484 a.u.), medium (middle, norm. intensity = 0.2429 a.u.) and high (right, norm. intensity = 0.6266 a.u.) bins showing intensity variation. **f.** The normalized intensity map divided into three bins namely, low [0 – 0.1], medium [0.1, 0.316], and high [0.316, 1] for the binning experiment. Capillaries from each bin (15, 10, and 10 from low, medium, and high, resp.) were extracted and sequences were identified (total number of capillaries scanned ≈ 92,000). **g.** DNA/RNA-sequencing read count data of the identified promoters from each bin. Two replicates of each are shown. **h.** Distribution of transcription rates of extracted promoters from bins. Cross represents mean in box-whisker plots.

Binning of clones is performed to reveal biological correlations between design and expression, with the transcription rate being the focus in this case[31]. We divided the normalized mRFP1 fluorescence intensity map into three logarithmically spaced bins: low, medium, and high (Fig. 3f). A representative capillary from each bin is depicted in Fig. 3e, showing pronounced inter-capillary variance in fluorescence signatures. Intra-capillary variance also exists, which is especially critical in recovering low-bin clones. While high-intensity capillaries have all their pixels above the threshold level, dim capillaries can have several pixels with intensities below this threshold (Supplementary Fig. 4a, c, f). Post-binarization, this latter case may result in capillaries being identified as clusters of several disjoint regions (see Supplementary Fig. 4b, d, g for comparison). Since each region is treated as a distinct entity on the intensity map, small regions can cause capillaries from higher bins to be misidentified as low-bin capillaries during sorting; that is, small regions can be false low-bin regions. To address this issue, we employed a pixel-dilation strategy (detailed in Methods) to eliminate false low-bin regions, which digitally coalesces small regions with the larger ones in capillaries (Supplementary Fig. 4e, h). The outcome is signified by a multifold reduction in the frequency of low-bin designations due to the elimination of false low-bin regions (Supplementary Fig. 4i). Hence, the misidentification of low-bin clones during binning was curtailed.

Fig. 3g illustrates the correlation between DNA and RNA sequencing read counts and fluorescence intensity for promoters randomly recovered from each bin. Clearly, RNA count is positively correlated with increasing fluorescence across low to high bins. Higher RNA counts are associated with a stronger fluorescence signal and, consequently, a higher protein titer. DNA counts also show a positive correlation, indicating that cell growth is crucial for protein titer. However, no clear trend in the transcription rate of promoters was observed to account for the DNA counts (Fig. 3h). The greater variance in the low and medium bin transcription rates reflects a wider range of promoters, including both strong and weak ones. This suggests that lower titer levels can result from weak promoters (low net RNA count) or toxic, strong promoters (low DNA count). We also link these observations (Fig. 3g, h) to inherent variations in protein expression (over 10×) among promoters with similar transcription rates[29]. Essentially, protein titer is closely linked to the influence of the promoter on cell growth and mRNA transcription. A promoter that maximizes mRNA translation across the colony—by optimizing the mRNA available per cell and the net cell count—is favorable for achieving high protein titers. Compared to screening the best performers, the binning algorithm required only a minimal additional time (101 seconds) for image processing, with recovery essentially being the rate-limiting step.

**mRNA Decay Effects on Protein Expression in 5' UTR Library**

Another critical component of the protein expression workflow is the 5' UTR of the mRNA. This region at the 5' end (Fig. 1b) directly impacts the decay of transcribed mRNAs within cells, thereby controlling the level of protein translation[32]. We utilized a library of 62,120 5' UTRs with varying exponential mRNA decay rates (detailed sequences published earlier[33]) and demonstrated the effect of mRNA half-lives[34] on protein titer through binning. This library of operons (Fig. 4a) encodes sfGFP and mRFP1 fluorescent reporters, with the 5' UTRs upstream of sfGFP acting as the control variable. Furthermore, using dual reporters enabled us to sort cells based on mathematical operations performed with distinct fluorescence signals to clarify the properties of RBS for each reporter.

Figures 4b and 4c display snapshots of sfGFP and mRFP1 fluorescence from the library in an array incubated at 37°C, respectively. Visually, these images reveal variation in the expression of both reporters and within each reporter. This variation is quantified in the normalized intensity maps (Fig. 4d) and raw intensity maps (Supplementary Fig. 5). While the range of sfGFP and mRFP1 intensities is similar, the frequency of low-intensity regions is significantly higher for mRFP1. We speculate that the lesser impact of highly unstable 5' UTRs on mRFP1 translation is the likely cause. We screened the library for sfGFP and mRFP1 signals separately, following the previously discussed protocol for binning. We scanned approximately 393,000 capillaries, compared to about 92,000 for the promoter library, to provide a broader representation of the larger 5' UTR library on the array. The mRNA half-life increases significantly as sfGFP fluorescence rises from low to high bins (Fig. 4e). 5' UTRs identified in the high bin exhibit on average over 4 times longer half-lives compared to those in the low bin, supporting sustained translation and leading to increased protein production and fluorescence intensity. Since 5' UTRs were designed solely to regulate sfGFP translation, with mRFP1 having an independent RBS, a similar effect on mRFP1 levels was not anticipated or observed (Fig. 4f). However, there is a positive correlation between the half-lives of sfGFP and mRFP1, which can be attributed to the operonic nature of the expression system.

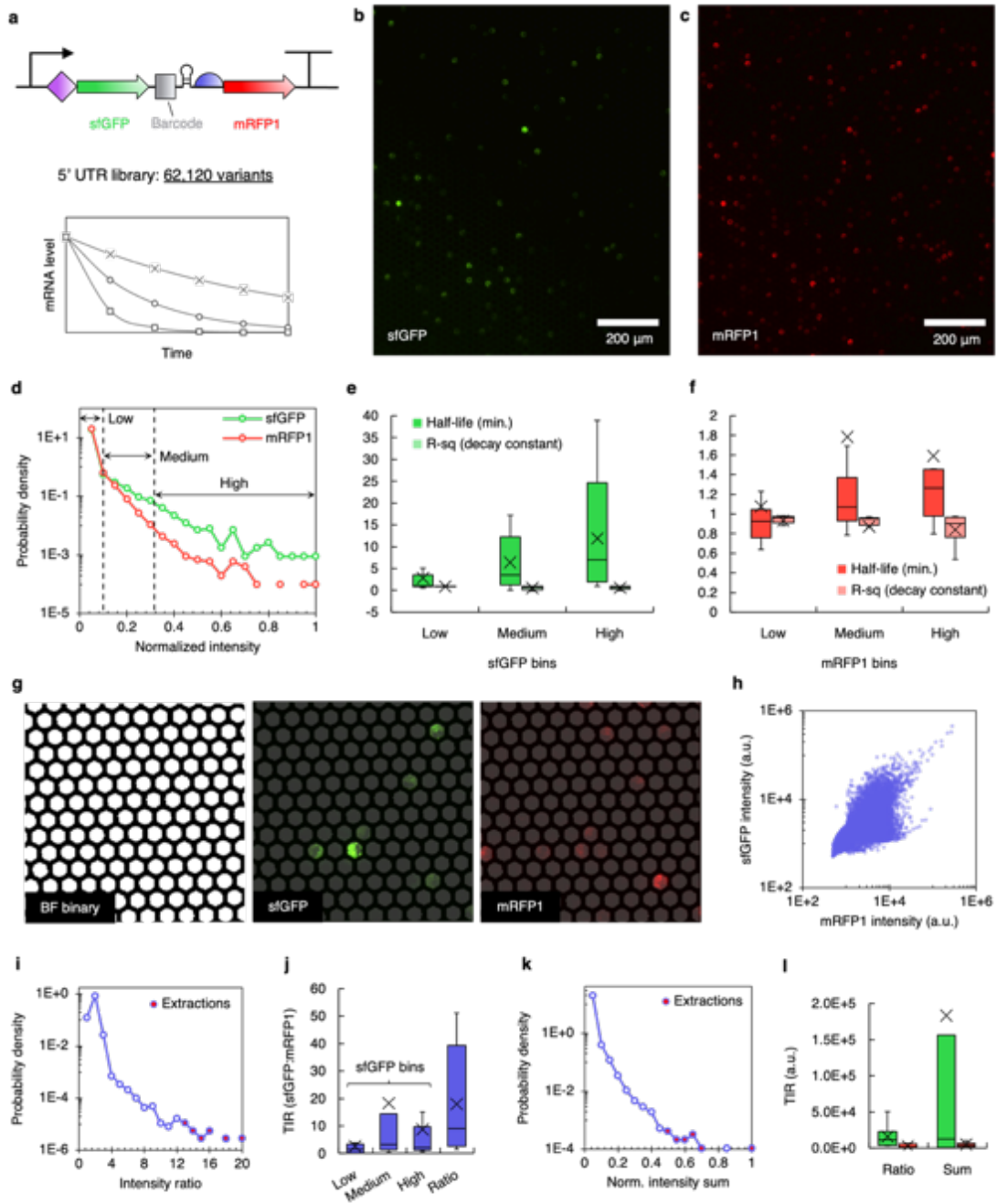

**Fig. 4. Screening of 5' UTR library. a.** Construction of the sfGFP and mRFP1-expressing plasmid is shown. The library consists of 62,120 5' UTRs (drawn in purple) with varying decay rates, with each assigned a unique barcode. **b, c.** Representative images of the array loaded with the UTR library after 12 hr. incubation period. sfGFP and mRFP1 signals at the same location on the arrays are shown. Brightness and contrast of images enhanced for visualization. **d.** Intensity maps for binning experiments for sfGFP and mRFP1 screening (total number of capillaries scanned ≈ 393,000; 10 extractions from each bin). **e, f.** Half-life and $R^2$ of fit for the decay constant are plotted for the identified 5' UTRs. **g.** Binarized brightfield image of an array as the mask for sfGFP and mRFP1 images is depicted. **h.** Scatter plot of sfGFP and mRFP1 fluorescence intensity in each capillary. **i.** Intensity map for the ratio-based (sfGFP/mRFP1; total number of capillaries scanned ≈ 364,000) screening is shown. Capillaries with non-

finite intensity ratios were not included in the intensity map. **j.** Ratios of translation initiation rates of RBSs for sfGFP to mRFP1 compared across sfGFP bins and ratio-based screened clones. **k.** Intensity map for the summation-based (sfGFP+mRFP1; total number of capillaries scanned ≈ 393,000) screening is depicted. Ten capillaries were extracted from the top bins and were sequence-identified in **(i)** and **(k)**. **l.** Translation initiation rates of sfGFP (in green) and mRFP1 (in red) RBS variants identified in the arithmetic-based screening experiments. Cross represents mean in box-whisker plots.

In addition to regulating protein expression through mRNA decay, the 5' UTR contains the ribosome binding site (RBS) for sfGFP, whose ribosome-recruiting strength directly influences translation. Each 5' UTR sequence yields a unique RBS with a specific translation initiation rate (TIR). The operonic nature of the plasmid necessitates relative analysis of reporter TIRs through dual reporter imaging. We screened capillaries based on fluorescence intensity ratios and the total sums of sfGFP and mRFP1. For the ratio experiment, we gathered sfGFP-mRFP1 fluorescence intensities in each capillary from discrete snapshots taken across the array. To process the sfGFP and mRFP1 signals in each capillary, the capillaries in fluorescence images of one reporter must be aligned with those of the other. Unlike image stitching, which can cause slight misalignment, discrete imaging avoids drift and ensures precise capillary tracking across the two sets of images (detailed in Supplementary Fig. 6). Additionally, we used brightfield snapshots as binary masks (Fig. 4g) for intensity quantification per capillary to eliminate errors arising from intra-capillary variations. While the discrete imaging protocol can be utilized for all mathematical operations, for the summation experiment, we captured the array with concurrent excitation of the reporters, making automated image stitching feasible. Both reporters were excited simultaneously using their respective illumination LEDs at the same irradiance level (illustrated in Supplementary Fig. 7). The remaining steps for screening capillaries exhibiting the highest intensity sums were identical to those for identifying the best-performing promoters.

Fig. 4h illustrates the fluorescence signatures of reporters within capillaries. The platform's ability to reveal highly differentiated cell populations indicated that 5' UTRs led to disproportionate protein expression by operons. Figs. 4i and 4k display nonlinearly decreasing profiles of intensity ratios and sums, with only a subset of clones maximizing each operation. The mRFP1 fluorescence in capillaries with high intensity ratios (sfGFP/mRFP1) was negligible (see Supplementary Fig. 8 for images). The ratio experiment resulted in the preferential recovery of clones with significantly greater relative TIRs for sfGFP (Fig. 4j) compared to those identified regardless of mRFP1 intensities (Fig. 4e). Thus, disproportionate sfGFP expression is associated with relatively stronger sfGFP RBSs compared to mRFP1 RBSs. The summation experiment, on the other hand, identified clones with higher absolute TIRs for both reporters on average, as compared to the ratio experiment (11.4 times for sfGFP and 1.7 times for mRFP1, Fig. 4l). Furthermore, the half-lives of these 5' UTRs were also greater (by 36.7% on average, Supplementary Fig. 9). These observations align with the respective screening modalities, such as disproportionate and cumulative expression. It is also worth noting that clones with extremely high sfGFP expression could potentially appear in both context experiments.

**Perturbation of Structural Proteins and Variability Analysis**

Producing recombinant proteins of SPs is notoriously challenging due to issues with folding, hydrophobicity, and post-translational modifications. SPs often possess complex shapes that are essential for their function. The cellular machinery of a host organism used for recombinant production may not be equipped to manage these intricate folding patterns. Likewise, the hydrophobic regions of SPs can lead to aggregates and inclusion bodies, which may be toxic to the host organism at high titers. Fig. 5a illustrates the design of the inducible plasmid featuring a T7 promoter and a fluorescent protein biosensor. We used the mCherry fluorescent reporter to quantify non-fluorescent SP titers in vivo. The four-nucleotide overlap (i.e., ATGA) between the SP stop codon and the mCherry start codon leads to ribosome re-initiation[35], which causes the translational coupling of mCherry and SP CDS. The additional experimental approaches discussed here include chemical perturbation of cells and measurement of growth indicators to analyze stochastic variability in recombinant gene expression. This platform is suitable for in-experiment induction, as the agarose gel can be exposed to chemicals (in this case, isopropyl β-D-1-thiogalactopyranoside (IPTG) as the inducer) that will diffuse through the matrix and into the capillaries over time (Fig. 5b). Instead of limiting the study to a single clone, we broadened our approach by using two proteins, namely cement and reflectin (see Supplementary Data 1 for amino acid sequences and CDSs), which have distinct hydrophobicity scores (Fig. 5a, calculated using models in ref.[36]). This property is associated with protein folding and aggregation[37].

To quantify the stochastic variability in SP production, we measured titer levels in cell cultures derived from single cells using a high-throughput approach. We anticipated the fluorescence signature from approximately 30% of the total capillaries; however, for SPs, we found that cells grew significantly in only 5.56 to 6.67% of the expected capillaries (i.e., only 500 to 600 out of roughly 9,000). This indicates that most cells in the starter culture were either nonviable or slow-growing (Supplementary Fig. 10a) and would not contribute to the amplification of the titer. Additionally, the distributions of capillary fluorescence (Supplementary Fig.11) spanned the entire intensity range, indicating that the expression was highly diffuse for both protein types (for comparison, see pFTV1 distribution in the inset of Supplementary Fig. 2b). The fluorescence snapshots in Fig. 5c and Supplementary Fig. 10 illustrate the differences in intensities of capillaries containing the two proteins. The plots in Fig. 5d display intensities (or titers) at various induction states after a 48-hour incubation. The basal titers for both proteins are higher than those at the respective induced state (t = 6 hr) due to the added toxicity and metabolic burden from the T7 promoter. However, compared to cement, reflectin exhibited significantly lower protein titers regardless of induction. While the reduced expression levels in reflectin can be linked to greater hydrophobicity and vice versa for cement, studying cell growth profiles across all conditions is essential to explain titer levels comprehensively.

We did not include magnetic beads in the single-clone experiments presented in Fig. 5. This removed optical interference from magnetic beads, ensuring that the degree of light transmission depended solely on cell count (Fig. 5e, f). Therefore, the extent of cell growth can be estimated by relatively quantifying absorbance using capillary intensity in brightfield snapshots. Fig. 5g (raw data in Supplementary Fig. 12c – h) displays the absorbance distribution per capillary above the baseline levels of empty capillaries. Regardless of induction, 95.5% more capillaries on average achieved absorbance over 0.2 a.u. with cement than with reflectin, which clusters at lower absorbance levels (87% of capillaries in the 0.1 a.u. – 0.2 a.u. range on average), indicating that reflectin expression significantly inhibited growth. This observation aligns with the lower mCherry fluorescence intensities associated with reflectin.

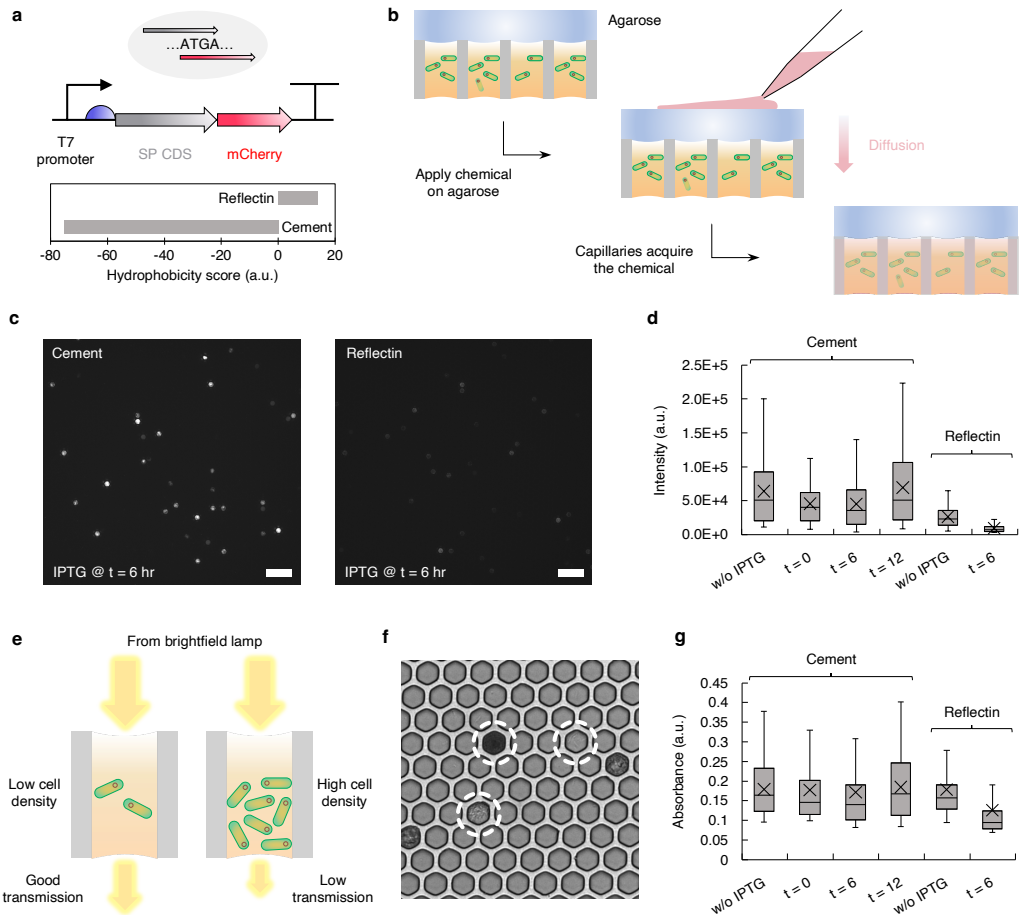

**Fig. 5. Chemical perturbation of structural proteins and screening. a.** Plasmid map with SP linked to mCherry fluorescent protein is shown. SP CDS is changed for cement and reflectin proteins accordingly, while keeping the RBS strength comparable. The hydrophobicity scores for cement (−75.26 a.u.) and reflectin (14 a.u.) are plotted. **b.** The strategy of chemical perturbation of cells is depicted. The concentrated chemical added to the agarose gel top face, which diffuses through the gel into the capillaries over time. **c.** Representative fluorescence images of cement and reflectin proteins with IPTG induction at t = 6 hr after 48 hr incubation period are shown. Scale bars = 100 μm. Brightness enhanced for visualization. **d.** The distribution of fluorescence intensity of capillaries (top 500) at various induction conditions after 48 hr. incubation period are plotted for both proteins (total capillaries scanned for each experiment ≈ 30,000). **e.** Schematic depicting the effect of cell density on absorption of brightfield light in capillaries. **f.** Representative brightfield image of cement protein array (at t = 6 hr induction) showing capillaries with varying levels of cell densities and absorbance. **g.** Brightfield absorbance distributions (top 500 capillaries) for cement and reflectin illustrating better cell-growth profiles of cement, and with delayed IPTG induction. Cross represents mean in box-whisker plots.

In addition to the hydrophobicity of amino acids, the timing of induction is crucial for cell growth. On average, the number of cement capillaries that achieved an absorbance of over 0.2 a.u. without induction and with a t = 12 hr induction is 42.3% higher than under early induction conditions. The variability in protein titer was confirmed through fluorescence assays of cement at these induction time points (t = 0, 6 hr, and 12 hr). Cement protein exhibited high net fluorescence intensity even without IPTG due to the basal expression of the T7 reporter (Fig. 5d). The intensity decreased significantly with early inductions at t = 0 and 6 hr. However, with a t = 12 hr induction, we reached a protein titer comparable to the non-induced state. Thus, the separation of growth and production phases was achieved with late-stage induction. This ensured that cells had ample time for growth (in the exponential growth phase) before induced expression, thereby mitigating metabolic burden with prior growth for improved titers.

**DISCUSSION**

We conducted high-throughput population screening utilizing fluorescent reporters on a microcapillary array platform. This platform enables the consistent growth of spatially isolated cells within capillary channels and allows for their precise recovery. In this study, we screened over $3.9 \times 10^5$ capillaries with a turnaround time of ten minutes to an hour. Microwell plate readers[14] and multi-parallel bioreactors[15,16,38] are alternative commercial technologies that allow for the screening of cell populations. However, these technologies have limited throughput, with the number of clones per screen restricted to $10^3$. Screening libraries on microwell plates is impractical, as it necessitates the differentiation of thousands of variants on agar plates, followed by picking and inoculation. The absence of an inherent mechanism for spatial isolation of cells renders these technologies unfit for library cloning protocols that yield mixtures of variants as final products. Microcapillary arrays offer screening rates that are $10^2$ to $10^3$ orders of magnitude higher, outperforming plate readers even in analyses of single variants. We demonstrated this by scanning approximately 183,000 SP capillaries (containing around 54,000 clones) per session, a capacity that can be extended as needed. Furthermore, the superior throughput of arrays helps reduce data noise by ensuring parallel replicates of variants.

Our platform ensures the accuracy of binning gates because, unlike FACS, these are determined after screening the entire population rather than just a representative sample. Additionally, we have customized the imaging and analysis protocols to facilitate the sorting of dual-reporter systems based on unique fluorescence signatures and specific mathematical functions. To our knowledge, such a modality in screening systems has yet to be introduced. Lastly, we have demonstrated in-experiment induction of SPs, with titer measurements in vivo and cell count estimates using brightfield absorbance.

We demonstrated the potential of arrays in optimizing biomanufacturability by screening libraries of promoters and 5' UTRs, which are among the most critical plasmid elements. After testing a range of transcription rates on the order of $10^6$, we found that promoters with moderate transcription rates were optimal for mRFP1 protein titer due to enhanced cell proliferation. Additionally, screening a library of 62,120 5' UTRs revealed a direct correlation between protein titer and mRNA half-life. Furthermore, we quantified the highly differentiated expression of sfGFP and mRFP1 in clones through dual-reporter imaging. We identified small subsets of clones that maximized the disproportionate sfGFP expression along with cumulative sfGFP-mRFP1 expression, which correlated with the respective RBS strength. Such features may contribute to fundamental studies involving operons, such as optimizing relative protein expression, increasing overall protein levels, and analyzing various elements like promoters and RBSs.

This study also focused on evaluating the variability in recombinant protein expression using SPs. We showed that the added toxicity of these proteins caused most cells in cultures to be either nonviable or to grow at reduced rates, resulting in wide titer distributions compared to the narrow distributions for mRFP1. The conventional strategy of in-experiment induction at various time points was replicated, demonstrating the separation of growth and production phases, which led to improved titers. Fluorescence imaging and direct measurements of cell growth in capillaries using brightfield absorbance supported these experiments, which had not been demonstrated with microcapillary arrays before. This study also confirmed conventional knowledge about expression systems, including the inhibited growth due to the T7 induction system, the impact of the hydrophobicity score of the amino acid sequence, and the evolution of culture properties over time. Additionally, a fluorescent biosensor was designed for in vivo measurements to eliminate the need for fluorescent tags and invasive protocols for SP concentration determination. The platform has the potential to aid in discovering novel protein sequences with greater biomanufacturability while preserving functionality, especially when complemented with mutagenesis and directed evolution[11,39].

Our platform facilitates the screening of various fluorescent reporters in the visible region (peak wavelengths from 405 nm to 635 nm, covering DAPI to Cy5), controls irradiance levels in 1% increments, allows for individual and simultaneous excitation, supports multiple image acquisition methods, and enables custom image analysis. We have demonstrated these features exclusively through the various experiments discussed above. The two main challenges of screening—namely, the creation of false low bin regions with low-fluorescing clones and drift during automated image acquisition—were addressed through image manipulation and discrete imaging, respectively. From a user-friendliness perspective, microscope environmental chambers[40] can help eliminate the need to transfer arrays to incubators between successive imaging sessions.

In summary, our microcapillary array platform enables rapid multiplex analyses of plasmid libraries, which would take weeks of continuous effort (such as laborious agar plating and colony picking) with conventional technologies. The relevance of miniature growth chambers in understanding microorganism behavior has been significant concern[41]. However, microcapillary arrays show expression-growth profiles that correlate with DNA/RNA sequencing markers derived from traditional culturing techniques. The spatial isolation of clones prevents biological crosstalk and allows users to study the unique phenotypic behaviors independently. Using this platform, we outlined the influence of several key plasmid components and demonstrated its potential to enhance the understanding of the complex phenomenon of protein expression.

## DATA AVAILABILITY

The authors declare that data supporting the findings of this study are available within the paper and its supplementary information files.

## AUTHOR CONTRIBUTIONS

M.C.D. conceived the project. H.M.S. provided the pFTV1, promoter library, and 5' UTR library. H.E.A. synthesized SP plasmids. T.M.B. and K.S. developed the platform. K.S. performed the screening experiments. K.S. wrote the manuscript together with M.C.D. All authors participated in manuscript revisions, discussion, and interpretation of the data.

## ACKNOWLDEGMENTS

The authors thank Penn State's Materials Research Institute and Huck Life Science Institute staff members for helpful discussions. We also thank William Humphries and William Hilinski of B&B Microscopes for their assistance.

## COMPETING INTERESTS

The authors declare that authors have issued and pending patents.

## ACKNOWLEDGEMENT OF SPONSORSHIP STATEMENT

This effort was sponsored in whole or in part by the Central Intelligence Agency (CIA), through CIA Federal Labs. The U.S. Government is authorized to reproduce and distribute reprints for Governmental purposes notwithstanding any copyright notation thereon.

## DISCLAIMER

The views and conclusions contained herein are those of the authors and should not be interpreted as necessarily representing the official policies or endorsements, either expressed or implied, of the Central Intelligence Agency.

## METHODS

**Instrumentation:** The platform (Fig. 2b) is based on an inverted fluorescence microscope (Olympus IX73) with a multi-channel LED illuminator (CoolLed pE-300$^{Ultra}$ or pE-400$^{Max}$). The microscope was equipped with a digital camera (Olympus DP23M), motorized XY stage (Marzhauser Wetzlar Tango 3), and motorized focus drive (Marzhauser Wetzlar). The proprietary software cellSens (Olympus) was used for imaging and stage control. The transmission characteristics of the filter cubes utilized (69302, 89401, Chroma) are illustrated in Supplementary Fig. 13. Microcapillary arrays with 20 μm capillary diameter (INCOM, Inc.) were used throughout all experiments, which contain about $8 \times 10^5$ capillaries in total. A 405 nm laser diode with a collimator was used for laser extraction. The laser was pulsed using a custom microcontroller. The pulse train was optimized (2 ms pulse separation, 2 ms pulse width, 4 – 5 in number at 118 DAQ level (power ≈ 75 mW); Supplementary Fig. 1c) to achieve the highest extraction efficiency. The power of the UV laser beam emanating from the objective was measured using an optical meter and sensor kit (Model 843-R, Newport Corp.). The laser was directed through the main objective by a beam reflector. The objective focused the collimated laser beam into a ≈ 10 μm spot, aligned with the face of the capillary to be extracted. Although pooled extraction has been demonstrated previously[23], we conducted manual extractions here to confirm the recovery of contents on slips after each iteration.

**Imaging and sorting:** All snapshots were acquired with a 10× objective (UPlanFL N, NA = 0.30, Olympus). An image of the loaded area of an array was constructed by stitching snapshots (3088 × 2076 pixels, about 1482 × 996 μm) together in an automated fashion (unless stated otherwise). Snapshots were obtained through raster scanning along a pre-defined path in a 2D plane (10,210 and 7,150 capillaries per second at 100 ms and 200 ms exposure, resp.), with the focus being automatically adjusted at each subsequent location. The generated image was processed with custom MATLAB code within $10^2$ seconds. The images were converted to grayscale and then binarized using the `imbinarize` function with the thresholding level defined by the `graythresh` function. Fluorescent capillaries were identified as white regions on a black background. Properties (centroid, area, and pixel indices) for all regions were obtained using the `regionprops` function. The fluorescence intensity of a region was calculated by summing the grayscale pixel values of all the comprising pixels. All regions (and their properties) were stored in a matrix and sorted for the measured intensity (or a mathematical operation). Before extraction, properties of regions of interest were analyzed, and care was taken to omit regions with areas deviating from normal substantially (e.g., coalesced regions). The coordinates of all places in the image were calibrated with the sample stage coordinates. Using the calibrated coordinates, regions of interest were swiftly aligned with the UV laser spot in cellSens for extraction. In all instances in the text, the total number of capillaries scanned refers to all the capillaries in the image frame (with or without cells).

**Microcapillary array loading, screening, and sequencing:** The arrays were sterilized before use. A mixture of magnetic beads (Dynabeads MyOne Silane, Invitrogen) and cells in media (Luria broth (LB)) was loaded onto the array. The small diameter makes it easier for liquid to pass through and be contained because of the surface tension of the liquid.[42]. This property is utilized to encapsulate cell cultures in capillaries, providing a controlled microenvironment for cell proliferation. To prevent capillaries from being overpopulated, we adjusted the cell concentration in the loading mixture to ensure that coverage remains below 33%. On average, one in every three capillaries acquires cells according to Poisson's statistics[23]. The magnetic beads cause a shift in the intensity distribution of capillaries; however, the distribution profile remains unchanged[23]. The concentration of magnetic beads was kept at 14 mg/ml. After loading, the array was overlaid with an agarose gel layer (2% w/v, 1 – 2 mm thick) and incubated in a sealed petri dish lined with moist wipes for a desired period. The setup of the sample stage is shown in Fig. 2b. The array rests on two standoffs on an indium titanium oxide (ITO) coated glass piece (Adafruit). The extraction slip (micro cover glasses, VWR) was inserted between the array and the glass piece. ITO glass was Joule-heated to 28 – 30 °C temperatures by passing 48 mA current using Bio-Rad PowerPac. This prevented condensation over the extraction slip during experiments. After screening, arrays were sterilized in 70% ethanol for several hours. Arrays were cleaned thoroughly under a DI water stream, with intermediate ultrasonication for 2 min.

PCRs for plasmid identification were conducted with Q5 high-fidelity kits (New England Biolabs). After each screening experiment, extracted contents were scraped off the extraction slip using a fine micropipette tip while applying 60 µl diluted Q5 reaction buffer (with 10× Q5 buffer to nuclease-free water ratio = 10 µl: 33.5 µl) on the slip and collected in an Eppendorf tube. PCR with 43.5 µl elution was carried out for 35 cycles at designated melting temperatures of primers and extension times for amplicons. PCR products were sequenced through the Oxford Nanopore technique (Plasmidsaurus), and alignment was done using the Context Aware AutoAlign algorithm (denovodna.com) for sequence identification.

**pFTV1 cell growth:** The plasmid design can be accessed at http://n2t.net/addgene:63848 (Addgene plasmid #63848, RRID: Addgene_63848). Isogenic colonies were picked and cultured in LB supplemented with chloramphenicol (20 µg/ml) for 5 hr at 37 °C and 250 RPM. The cell culture was diluted in fresh media (5 µl in 1 ml), and 10 µl was loaded onto an array. The array was incubated at 37 °C. Brightfield, and mRFP1 fluorescence snapshots of several array locations were acquired at t = 1 hr, 4 hr, 8 hr, and 24 hr incubation periods. After each round of imaging, the array was incubated, and the exact locations were imaged in subsequent rounds to record cell growth and protein expression. Brightfield snapshots were binarized and used as masks for fluorescence snapshots to calculate intensity per capillary.

**Binning experiments:** Fluorescence regions identified by image processing were grouped into three logarithmically spaced bins for normalized intensity (low (0 – 0.1], medium [0.1, 0.316], and high [0.316, 1]).

$$Normalized\ intensity = \frac{Intensity - \min{(Intensity)}}{\max(Intensity) - \min(Intensity)} \qquad (1)$$

Each bin's desired capillaries were extracted on a clean extraction slip. The original fluorescence image was used to screen directly for medium and high bins. However, for the low bin, the grayscale fluorescence image was dilated once before image processing using the `imdilate` function with a disc-shaped structural element of size 6. A new normalized intensity map was generated in which low bin capillaries were identified and extracted in desired numbers. Capillaries to be extracted from each bin were selected from the corresponding list randomly using the `randperm` function in MATLAB.

**Promoter library preparation and screening:** The detailed methodology of preparation of the promoter library is mentioned in ref.[29]. Cells were picked from the cryostock using a micropipette tip. Cells were cultured in LB supplemented with chloramphenicol (50 µg/ml) at 37 °C and 250 RPM until an $OD_{600}$ of 0.1 – 0.2 was reached. The cell culture was diluted in 100 µl fresh media

(as per relation: 2 µl starter culture at $OD_{600} = 0.1$) containing magnetic beads at 14 mg/ml conc. The diluted culture was loaded onto an array and incubated at 37 °C overnight. The extraction of top performers (3 replicates) and binning were performed with the library. A separate array was used for each experiment.

**5' UTR library preparation and screening:** The detailed methodology of preparation of the promoter library is mentioned in ref.[33]. The same protocol of array loading and incubation as for the promoter library was followed. Cells picked from the cryostock were used to prepare cell cultures. The volume of diluted culture loaded onto the array was increased to achieve greater array coverage. Binning of sfGFP and mRFP1 each and arithmetic-based screening were performed with the library. A separate array was used for each experiment. TIRs were calculated using the RBS calculator (denovodna.com), and the sequences are mentioned in Supplementary Data 1.

**Arithmetic-based screening:** Schematic in Supplementary Fig. 6 illustrates the algorithm for ratio-based screening. At several array locations, brightfield and fluorescence snapshots were acquired discretely. Brightfield snapshots were binarized, and resulting snapshots were used as binary masks for fluorescence snapshots. The mathematical operation was done on sfGFP and mRFP1 intensities for each capillary, and the cells with the greatest sfGFP to mRFP1 ratios were extracted, and the sequences of 5' UTRs were identified.

The previous approach of raster scanning and image stitching (as for binning) was followed for the summation-based screening experiment. Our fluorescence illumination system allowed for the use of multiple LEDs simultaneously for concurrent excitation of sfGFP and mRFP1. In systems where various LEDs cannot be used simultaneously, the ratio-based screening protocol can be used while switching the mathematical operation to summation.

**Structural proteins cloning and screening:** The plasmids were designed using the Genetic Systems Builder (all calculators are available at denovodna.com). The algorithm first optimizes DNA sequence for efficient expression of the gene of interest. All plasmids are transformed into a pool of ssDNA oligos which was sourced commercially (Twist Bioscience). Furthermore, PCR amplification primers for each oligo type were ordered separately. Using the specific primer pair, oligos of a desired plasmid were amplified and converted into dsDNA fragments through PCR. This PCR product was then purified and used for standard GG assembly. The assembled plasmids were transformed into *E. Coli* BL21-DE3 strain (NEB) using the standard protocol.

Isogenic colonies of SP proteins were inoculated in LB supplemented with chloramphenicol at 28 °C overnight, and the culture was utilized to load the arrays. The culture was diluted according to the relationship mentioned earlier for libraries. The arrays were incubated at 30 °C. For t = 0 induction, IPTG was added to the culture before loading at a final concentration of 400 µM. For in-experiment inductions, 0.1 mM IPTG stock was applied to the agarose gel surface and spread gently. The volume of the gel and loaded area of the array were considered to estimate the volume of the stock solution to be added to achieve a final concentration of 400 µM in capillaries. Diffusion is a kinetic process, and we attempted to measure the diffusion rate of IPTG into the capillaries through Fourier Transform Infrared Spectroscopy of the elution from arrays after several hours. However, the IPTG concentration was too low to be detected. Earlier studies on diffusion of similar molecules in agarose[43] suggest that saturation in the setup (through 1.5% w/v gel) would take 2 – 3 hours at 25 °C. For focused studies, dedicated diffusion analyses can be carried out for the molecule in question. A separate array was used for each experiment with an SP.

The discrete imaging technique was followed for screening SPs to minimize errors in fluorescence intensity measurements per capillary. For background correction, fluorescence snapshots of blank regions of arrays under the gel were acquired. After segmentation, brightfield images were acquired to act as binary masks for fluorescence images. For brightfield absorbance measurements, the irradiance of the light was reduced such that capillaries with low cell count posed significant obstruction for their detection. The brightfield images were segmented and used as binary masks to calculate brightfield intensity per capillary. Baseline correction of the distributions was performed manually using levels centered around the mode of each trend, as detailed in Supplementary Fig. 14. Finally, absorbance per capillary was calculated using the following relation:

$$Absorbance = -log_{10} \frac{Intensity}{Baseline\ Intensity} \quad (2)$$

It is to be noted that baseline correction does not alter the scattered distribution patterns of data points and only influences numerical values of absorbance. The capillaries were filtered for size, and the ones with face areas smaller than average were excluded from the analysis.